\documentclass[]{article}

\usepackage{etex}
\usepackage{hyperref}
\usepackage{tabularx}
\usepackage{bussproofs}
\usepackage{proof}
\usepackage{xypic}
\usepackage{fancyhdr}
\usepackage{apacite}

\usepackage[english]{babel}
\usepackage{url}
\usepackage{amsmath} 
\usepackage{amssymb}
\usepackage{color}

\newtheorem{definition}{Definition}
\newtheorem{theorem}{Theorem}

\newtheorem{proposition}{Proposition}

\newcommand{\alc}{\ensuremath{\mathcal{ALC}}}
\newcommand{\ialc}{\ensuremath{i\mathcal{ALC}}}
\newcommand{\system}[2]{\ensuremath{\textsf{#1}_{#2}}}
\newcommand{\nominal}[2]{\ensuremath{#1\colon{}\mkern-4mu #2}}

\newcommand{\SEQ}{\ensuremath{\Rightarrow}}

\newcommand{\dland}{\sqcap}        % concept intersection
\newcommand{\dlor}{\sqcup}         % concept union
\newcommand{\subs}{\sqsubseteq}    % concept subsumption
\def\fCenter{ \SEQ\ }

\newenvironment{dedsystem}
{
 \newcolumntype{Y}{>{\centering\arraybackslash}X}
 \setlength{\extrarowheight}{2.5ex}
 \tabularx{\textwidth}{YcY}}
{\endtabularx}

\title{On How Kelsenian Jurisprudence and Intuitionistic Logic help to  avoid Contrary-to-Duty paradoxes in Legal Ontologies}

\author{Edward Hermann Haeusler$^{a}$$^{\ast}$\thanks{$^{\ast}$Acknowledge CNPQ, prof. David Pearce, Universidad Polit\'{e}cnica de Madrid.} and Alexandre Rademaker$^{b}$ \\\vspace{6pt} 
  $^{a}${\em Department of Informatics, PUC-Rio }\\\vspace{6pt}
  $^{b}${\em IBM Research, Rio de Janeiro, Brazil}}

\begin{document}

\maketitle

\begin{abstract} 
In this article we show how Hans Kelsen jurisprudence and Intuitionistic logic are used to 
avoid the well-known contrary-to-duty (CTD) paradoxes, such as Chisholm paradoxes and its variants. 
This article uses an intuitionistic version of the ALC description logic, named iALC, to show how 
an ontology based on individually valid legal statements is able to avoid CTDs by providing models to them.
\end{abstract}

\section{Introduction}

Prof. Luis Fari\~{n}as del Cerro wrote a bunch of articles reporting the results he obtained by designing logics for very interesting and specific purposes. The elegance of the underlying ideas and the presentation form is out of discussion. Many researchers in logic would like to have his ability to extract logic from facts and their relationships, building new judgments. One of the authors of this article ever tried to be able to have this resulting research. When we were invited to contribute to prof. Luis Fari\~{n}as del Cerro Festschrifft this become the opportunity to report to the master maybe the only research that we conduct on the lines of defining a logic for a specific formalization. In this article, we report our results in the last seven years in the designing logics for legal ontologies. 

Classical First Order Logic has been widely used as a basis for ontology creation and reasoning in many domains. These domains naturally include Legal Knowledge and Jurisprudence. As we expect, consistency is an important issue for legal ontologies. However, due to their inherently normative feature, coherence (consistency) in legal ontologies is more subtle than in other domains. Consistency, or absence of logical contradictions, seems more difficult to maintain when more than one law system can judge a case, what we call a conflict of laws. There are some legal mechanisms to solve these conflicts such as stating privileged fori or other ruling jurisdiction. In most of the cases, the conflict is solved by adopting a law hierarchy or precedence, rather better, ordering on laws. Even under these precedence mechanisms, coherence is still a major issue in legal systems. Each layer in this legal hierarchy has to be consistent. Since consistency is a direct consequence of how one deals with the logical negation, negation is also a main concern in legal systems. Deontic Logic, here considered as an extension of Classical Logic, has been widely used to formalize the normative aspects of the legal knowledge. There is some disagreement on using deontic logic, and any of its variants, to this task. Since a seminal paper by Alchourron and Martino~\cite{AlchourronMartino1990}, the propositional aspect of laws has been under discussion. In~\cite{AlchourronMartino1990}, the authors argue that {\em laws} are not to be considered as propositions, in full agreement with Hans Kelsen jurisprudence. The Kelsenian approach to Legal Ontologies considers the term ``ontologies on laws'' more appropriate than ``law ontology''.  In previous works, we showed that Classical logic is not adequate to cope with a Kelsenian based Legal Ontology. Because of the popularity of Description Logic for expressing ontologies nowadays, we developed an Intuitionistic version of Description Logic particularly devised to express Legal Ontologies. This logic is called iALC. In this article, we show how iALC avoids some Contrary-to-duty paradoxes, as {\em Chisholm}  paradox and other paradoxes that appear in deontic logic, such as {\em the good samaritan} and {\em the knower}. For these paradoxes, we provide iALC models. Finally, we discuss the main role of the intuitionistic negation in this issue, finding out that its success may be a consequence of its paracomplete logical aspect. This investigation opens the use of other paracomplete logics in accomplishing a logical basis for Kelsenian legal ontologies, as a complementary solution to those based on paraconsistent logics, see~\cite{ConiglioNewton}. 

%%% Local Variables:
%%% mode: latex
%%% TeX-master: "CD-Paradoxes-iALC"
%%% End:

\section[Kelsenian Jurisprudence]{A brief discussion on Kelsenian Jurisprudence and its logic}
\label{jurisprudence}
A very important task in jurisprudence (legal theory) is to make precise the use of the term ``law'', the {\em individuation problem}, and it is one of the most fundamental open questions in jurisprudence. It requires firstly answering the question ``What is to count as one complete law?'' \cite{Raz1972}. There are two main approaches to answer this question. One approach is to consider ``the law'' as the result of a natural process that yields a set of norms responsible for stating perfect social behavior. Another approach is to consider ``the law'' as a set of individual legal statements, each of them created to enforce a positively desired behavior in the society. As a consequence, in the first approach, the norms say what are the best morally speaking accepted state of affairs in a particular society, while, in the second approach, each legal statement rules an aspect of the society that the legislature wants to enforce the behavior. The first is more related to what is called {\em Natural Law} and the last to {\em Legal positivism}. We can say that the {\em Legal positivism} is closer to the way modeling is taken in Computer Science. In the {\em natural} approach to the law, it is even harder to define a system of laws than in the {\em legal positivism}. The natural approach demands stronger knowledge of the interdependency between the underlying legal statements than legal positivism. Because of that, the {\em natural} approach, in essence, is harder to be shared with practical jurisprudence principles, since they firstly are concerned to justify the law, on an essentially moral basis. This justification is quite hard to maintain from a practical point of view.

The coherence of ``the law'' in both approaches is essential. A debate on whether coherence is built-in by the restrictions induced by Nature in an evolutionary way, or whether coherence should be an object of knowledge management, seems to be a long debate. Despite that, legal positivism seems to be more suitable to Legal Artificial Intelligence. From the logical point of view, the {\em natural} approach is also harder to deal with than the {\em positivist} one. When describing a morally desired state-of-affairs, the logical statements take the form of propositions that has as a model best of the moral worlds. Deontic logic is suitable to be used to fulfill this task. However, a legal statement (``a law'') is essentially an individual sentence that can also be seen as an order (mandatory command), and hence, it is not a proposition at all. As a consequence, deontic logic is not appropriate to be used in knowledge bases. Besides that,   \cite{Valente1995} shows that deontic logic does not properly distinguish between the normative status of a situation from the normative status of a norm (rule). We think that the best jurisprudence basis for Legal ontologies and reasoning is {\em Legal positivism}. Thus, we will be talking a legal ontology as an ontology about (individual) laws, and not an ontology on ``the law''.  

Hans Kelsen initialized the {\em Legal positivism} tradition in 1934, for a contemporary reference see \cite{Kelsen1991}.  He used this positive aspect of the legislature to define a theory of {\em pure law} and applied it to the problem of transfer citizen's rights and obligations from one country to other when crossing boarders. He produces a quite good understanding of what nowadays we denominate {\bf Private International Law}.  This achievement was so important that in many references on international law, Kelsen jurisprudence is the basis for discussions on conflict-of-laws derived from different statements coming from different {\em fori}.~\footnote{``It is one of Kelsen's frequently repeated doctrines that conflict of norms, in the absence of a normative procedure for resolving the conflict, shatters the concept of a unified system'', is highly emphasized in Hughes~\cite{GrahamHughes1971}, for example, and it is one of the principles most cited when Kelsen jurisprudence is presented.}

In what follows we introduce the main terminology and concepts of Kelsenian jurisprudence that we use in this article. We can summarize Kelsen theory of pure law in three principles:

\begin{enumerate}
\item\label{I} According to what was discussed above, individually valid legal statements are the first-class citizens of our ontology. Thus, only inhabitants of the Legal knowledge base are individual laws, see~\cite[supra note 5, pp 9-10]{KelsenRRII} ~\footnote{Kelsen takes norms and valid norms as synonyms. To say that a legal norm is valid is to say that it exists, is affirmed by Kelsen}. For example, if it is the case that Maria is married with John, and, this was legally celebrated, then ``Maria-married-with-John'' is an individually valid legal statement, and hence, it is a member of the Legal Ontology;

\item\label{II} Kelsen also says that that the validity of a legal norm can only be provided concerning the validity of another, and higher,  one. So, $n_1$, a norm, is legally valid if, and only if, it was created or promulgated in agreement with other, and higher, legally valid norm, $n_2$. This justification induces a precedence relationship between norms that is transitive, that is, if $n_1$ precedes $n_2$, and, $n_2$ precedes $n_3$, then $n_1$ recedes $n_3$;~\footnote{See ~\cite[supra note 5, p. 196-7]{KelsenRRII}. This can be also found in~\cite[supra note 5, p. 110-1]{KelsenGTLS}}

\item\label{III} There is a mechanism for relating laws from one Legal system to another, the so-called ``choice-of-law rule''. This mechanism is very important to the development of a concept of International Law. Assume that {\em Mary-is-married-with-John} is an individual legal statement in legal system $A$. Assume also that Mary is a citizen of a country adopting legal system $B$. Is there any legal statement in $B$ ensuring that Mary is married in $B$? Well, this depends on $B$ itself, but there is a way to connected the individual law {\em Mary-is-married-with-John} in $A$ to {\em Mary-is-married-with-John} in $B$. In some legal systems, this is accomplished by what Kelsen denominated ``a connection''. As shown in the following quotation from~\cite[page 247]{KelsenGTLS}, the connection between the laws of $A$ and $B$ is made by reference, but, in fact, each law belongs to its respective legal system. In this specific case we can consider {\em Mary-is-married-with-John} in system $A$ is connected to {\em Mary-is-married-with-John} in legal system $B$ the connection {\em Lex Loci Celebrationis}.
\begin{quote}
... the law of one State prescribes the application of the law of another State, and the latter does not object or demand it. It has no right to do so since it is not really its own law which is applied by the other State. The latter applies norms of its own law. The fact that these norms have the same contents as corresponding norms of another State does not concern the latter...Since the specific technique of these norms consists in ``referring'' to the norms of another system and by so doing incorporating norms of identical contents into their own legal system, it would be more justifiable to call them ``reference rules''... The reference rule, that is ... the norm regulating the application of foreign law, may be distinguished from the norm to be applied, that is, the norm referred to. Only the former is a norm of private international law. But from a functional point of view, the one is essentially connected with the other.
\end{quote}
\end{enumerate}

Nowadays it is a common terminology in Private International Law the use of the connecting factors or legal connections between individual laws in a different legal system. Only to enumerate some of them: {\em Lex-Domicilii}, {\em Lex-Patriae}, {\em Lex-loci-contratum}, {\em Lex-loci-solutionis}, etc.

There is a philosophical problem with the principle~\ref{II} above. It demands the existence of basic laws. These basic laws do not have their validity/existence as a consequence of other more basic laws. Kelsen name these basic laws {\em Grundnorms}. Their validity is based on legislature acts and in a certain sense is derived from the sovereign of the State. It is out of the scope of this article to discuss such problem in Kelsen's jurisprudence. We take as granted that Kelsen jurisprudence can adequately support most of the existent legal systems, a definitively not an unreal working hypothesis. 

From the three principles above, we have some very simple ontological {\bf commitments}:

\begin{description}
\item[I]\label{commitI} Individuals are laws;
\item[II]\label{commitII} There is a transitive and reflexive
  relationship between individual laws that reflects the natural
  precedence relationship between laws;
\item[III]\label{commitIII} There are legal connections between
  individual laws in different legal systems or between different {\em
    fori} in the same broader legal system.
\end{description}

From these commitments, we derive the basic constructs of the logic \ialc. In the first place, our legal ontology relates concepts to legal systems. Description logics uses nominals to refer to individuals. So, an expression as $i:A$, stands for $i$ is an individual law, belonging to the legal system $A$, a concept. 

From commitment~\ref{commitII} we consider an expression as $i\preceq j$ standing for the individual law $i$ legally precedes individual law $j$. The subsumption relationship $A\subseteq B$, from description logic, denotes that $A$ is a legal subsystem of $B$. One could interpret this relation as the inclusion relationship.~\footnote{In Classical ALC this is just the case, but we shown here that classical reasoning it is not a good choice for dealing with legal ontologies} We discuss the implications of using negated contents together with Kelsenian jurisprudence in the following. This can be found in~\cite{HPR:2011,HPR:2010b,HPR:2010a,HPR:2010ab} too. 

Under the classical setting, a negated concept $\neg A$ denotes the {\bf set} of all inhabitants of the domain that do not belong to the interpretation of $A$. Under ontological commitment~\ref{commitI} there is no individual law that does not exist in, belong to, the domain. Since norms and laws are not propositions, it is a complete nonsense to negate a law. As we already seen, we can negate a concept on laws. Consider the collection of all Brazilian individual laws. Call it $BR$. In a classical setting $BR\dlor \neg BR$ is the universe of laws. Thus, any law that it is not in $BR$ has to be a law outside $BR$, that is, belonging to $\neg BR$. For example, if Peter is 17 years old, it is not liable according to the Brazilian law. Is \textsc{PeterIsLiable} a valid law at all? If so, it has to belong to $\neg BR$. Using Kelsen in a classical setting, individual laws not belonging to a concept automatically belong to its complementary concept. The problem with this is that it is possible to create laws outside a jurisdiction or forum by the very simple act of considering or experimenting a legal situation. Nowadays in Brazil, the parliament is discussing the liability under the 16 years. By the simple fact of discussing the validity of their corresponding individual laws, we are forced to accept they exist outside the Brazilian legal system. We do not consider this feature appropriate to legal ontology definition. Dealing with negations every time we assume the existence of a law may bring unnecessary complexity to legal ontology definition. Because the precedence relationship between laws, cf. ontological commitment~\ref{commitII}, there is a natural alternative to classical logic, the intuitionistic logic (IL). According to  IL semantics, $i:\neg A$, iff, for each law $j$, such that $i\preceq j$, it is not the case that $j: A$. This semantics means that $i$ does not provide any legal support for any individual law belong to $A$, which agrees with Kelsen jurisprudence on the hierarchy of individual laws. 

Commitment~\ref{commitIII} gives rise to expressions of the form $m$ \textsc{LexLociCelebrationis} $m$, where $m$ is \textsc{MaryIsMarriedWithJohn} and \textsc{LexLociCelebrationis} is a  legal connection. Thus, if $Abroad$ is the concept that represents all laws in Portugal, then the concept $\exists\;\textsc{LexLociCelebrationis}\;Portugal$ represents the Brazilian individual laws stating that Portuguese marriage is valid in Brazil. The private international law of any country is a collection of laws stated in similar ways for every possible legal connection. In~\cite{HPR:2010b} it is shown in detail a judicial case deriving that a renting contract is solving a conflict of laws in space through private international law. 

%%% Local Variables:
%%% mode: latex
%%% TeX-master: "CD-Paradoxes-iALC"
%%% End:

\section[Ontological Criteria]{Some philosophical discussion on the ontological criteria taken on using Kelsen in legal ontologies} 

We base our work on two ontological criteria:~\footnote{see Quine's ``On What there is'' article and \url{http://plato.stanford.edu/entries/simplicity}, for example to a primer ontological criteria} 1- Ontological Commitment (due to W.Quine), our logical approach is ontologically committed to Valid Legal Statements only, as discussed in section~\ref{jurisprudence}. The only nominals occurring in our logic language are valid individual laws, and; 2-Ontological Parsimony, which is strongly related to Quine's ontological commitment too, with a mention of its stronger version also known as Occam's Razor, here denoted as OR. The second criteria is based on: ``One `easy' case where OR can be straightforwardly applied is when a theory T, postulates entities which are explanatorily idle. Excising these entities from T produces a second theory, T*, which has the same theoretical virtues as T but a smaller set of ontological commitments. Hence, according to OR, it is rational to pick T* over T.''

We observe that nominals, representing individuals, denote only valid individual laws and nothing in the \ialc\ language described in the following section,  is committed with non-valid individual laws, according to the second ontological criterion above,  we do not have to consider non-valid individual laws. Technically speaking there is no element in the \ialc\ language able to denote an invalid individual law in any model of  any  \ialc\ theory. If something is a valid individual law regarded some legal system in some place in the world, then this individual  belongs to our semantic universe. 

This philosophical basis allows us to have only sets of valid individuals as semantics for \ialc\ theories. Thus, as the a reviewer have already observed, this implies that  $\neg A$ is the set of individual laws  holding outside Brazil, and the classical negation is not adequate to denote this set. If we get $\neg A$ meaning ``individual laws that \emph{do not} hold in Brazil'', the set of laws being a proper subset of the universe, and $A$ is the conjunctive property ``laws + holds in $A$''. Then the complement, $\neg A$ would be all elements of the universe which are either not a valid individual law or do not hold in Brazil. But there is no way to take the semantics in this way, for the semantics we get from the ontological commitment~\ref{commitI} from section~\ref{jurisprudence} is given by ``The individual valid laws holding outside of Brazil''.

Finally, concerning contradictory individual laws, they can coexist in the same universe, since they are there because they hold in distinct legal systems. In fact they are apparently contradictory. For example, ``There is death penalty'' and ``Death sentence is not allowed'' can coexist, since there are countries where each of these legal statement are valid. Concretely: ``There is death penalty'':Iran and ``Death sentence is not allowed'':Brazil.

%%% Local Variables:
%%% mode: latex
%%% TeX-master: "CD-Paradoxes-iALC"
%%% End:

\section{The Logic \ialc}\label{sec:intro-ialc}

Classical Description Logic has been widely used as a basis for
ontology creation and reasoning in many knowledge specific domains,
including Legal AI.
% These specific domains naturally include Legal AI.
%% As in any other domain, consistency is an important issue for legal
%% ontologies.  However,
%% Due to the apparent inherently normative feature of Legal ontologies,
%% coherence (consistency) in legal ontologies is more subtle than in
%% other domains. Consistency, or absence of logical contradictions,
%% seems harder to maintain when more than one {\em law system}
%% can judge a case. This is called a {\em conflict of laws}. There are
%% some legal mechanisms to solve these conflicts, some of them stating
%% privileged fori, other ruling jurisdiction, etc. In most of the cases,
%% the conflict is solved by admiting a law hierarchy or a law
%% precedence. Even using these mechanisms, coherence is still a major
%% issue in legal systems since each layer in this legal hierarchy has to
%% be consistent.  As consistency is a direct consequence of how one
%% deals with logical negation and subsumption. 
%% % , negation is also a main concern of legal systems.
%% Negation and subsumption play a central role in ontology coherence. 

An adequate intuitionistic semantics for negation in a legal domain
comes to the fore when we take legally valid individual statements as
the inhabitants of our legal ontology. This allows us to elegantly
deal with particular situations of legal coherence, such as conflict
of laws, as those solved by Private International Law analysis. In
\cite{HPR:2010ab,HPR:2010b,HPR:2010a} we present an Intuitionistic
Description Logic, called \ialc\ for Intuitionistic \alc\ (for
Attributive Language with Complements, the canonical classical
description logic system). A labeled sequent calculus for \ialc\,
based on a labeled sequent calculus for \alc~\cite{Rademaker2012}, was
also presented. In these previous articles, we discussed the
jurisprudence foundation of our system, and show how we can perform a
coherence analysis of ``Conflict of Laws in Space'' by means of
\ialc. This conflict happens when several laws can be applied, with
different outcomes, to a case depending on the place where the case
occurs. Typical examples are those ruling the rights of a citizen
abroad.

In \cite{HPR:2010b}, the semantics of \ialc\ is precisely provided and  follows the framework
for constructive modal logics presented by Simpson~\cite{simpson95}
and adapted to description languages by Paiva~\cite{depaiva2003}. In the cited reference, we 
applied \ialc\  to the problem of formalizing legal knowledge. 

Description Logics are an important knowledge representation
formalism, unifying and giving a logical basis to the well known AI
frame-based systems of the eighties. Description logics are very
popular right now. Given the existent and proposed applications of the
Semantic Web, there has been a fair amount of work into finding the
most well-behaved system of description logic that has the broadest
application, for any specific domain. Description logics tend to come
in families of logical systems, depending on which concept
constructors you allow in the logic. Since description logics came
into existence as fragments of first-order logic chosen to find the
best trade-off possible between expressiveness and tractability of the
fragment, several systems were discussed and in the taxonomy of
systems that emerged the \alc\ has come to be known as the canonical
one. The basic building blocks of description logics are {\em
  concepts}, {\em roles} and {\em individuals}. Concepts are described as
unary predicates in usual first-order logic and roles as binary atomic
predicates used to modify the concepts.

As discussed in \cite{depaiva2003}, considering versions of {\em
  constructive} description logics makes sense, both from a
theoretical and from a practical viewpoint.  There are several
possible and sensible ways of defining {\em constructive} description
logics, whether your motivation is natural language semantics,
\cite{depaiva2003}, or Legal AI, \cite{HPR:2010ab}. As far as
{\em constructive} description logics are concerned, Mendler and
Scheele have worked out a very compelling system $c{\cal
  ALC}$~\cite{mendler-scheele}, based on the constructive modal logic
\textsf{CK}~\cite{bellinetal}), one possible choice %\footnote{This system has
%  categorical semantics, which are not very easy to obtain for modal
%  logics.}
 for us. However in this note we follow a different
path and describe a constructive version of \alc, based on the
framework for constructive modal logics developed by Simpson (the
system \textsf{IK}) in his phd thesis~\cite{simpson95}
% We call our system \ialc\ for Intuitionistic \alc
(For a proof-theoretic comparison between the constructive modal
logics $\sf CK$ and $\sf IK$ one can see ~\cite{ranalter09}).

Our motivation, besides Simpson's work, is the framework developed by
Bra\"{u}ner and de Paiva in \cite{brauner-depaiva2006} for
constructive Hybrid Logics. We reason that having already frameworks
for constructive modal and constructive hybrid logics in the labelled
style of Simpson, we might end up with the best style of constructive
description logics, in terms of both solid foundations and ease of
implementation. Since submitting this paper we have been told about
the master thesis of Cl\'ement~\cite{clement08} which follows broadly
similar lines. Cl\'ement proves soundness and completeness of this 
system and then provides a focused version of it, a very interesting development, as focused systems are,
apparently, very useful for proof search.

Building up from the Simpson's constructive modal logics (called here
IML), in \cite{brauner-depaiva2006}, it is introduced
\emph{intuitionistic hybrid logics}, denoted by \textsf{IHL}. Hybrid
logics add to usual modal logics a new kind of propositional symbols,
the \emph{nominals}, and also the so-called \emph{satisfaction
  operators}. A nominal is assumed to be true at exactly one world, so
a nominal can be considered the name of a world. If $x$ is a nominal
and $X$ is an arbitrary formula, then a new formula \nominal{x}{X}
called a satisfaction statement can be formed. The satisfaction
statement \nominal{x}{X} expresses that the formula $X$ is true at one
particular world, namely the world denoted by $x$. In hindsight one
can see that IML shares with hybrid formalisms the idea of making the
possible-world semantics part of the deductive system. While IML makes
the relationship between worlds (e.g., $xRy$) part of the deductive
system, IHL goes one step further and sees the worlds themselves $x,y$
as part of the deductive system, (as they are now nominals) and the
satisfaction relation itself as part of the deductive system, as it is
now a syntactic operator, with modality-like properties.

Our Sequent Calculus for \ialc\ was first presented in \cite{PHR:2010}
where we briefly described the immediate properties of this system and
most importantly we discuss a case study of the use of \ialc\ in legal
AI.

A very importante obsertation is that this article corrects and extends the presentation of \ialc\ appearing
in all previous articles. It points out the difference between \ialc\
and the intuitionistic hybrid logic presented in
\cite{depaiva2003}. Completeness and soundness proofs are revised. A
discussion on the computational complexity of \ialc\ is also taken.

\section{Intuitionistic ALC}\label{sec:ialc}

The \ialc\ logic is based on the framework for intuitionistic modal
logic IK proposed in \cite{simpson95,Fisher1984,plotkin1986}. These
modal logics arise from interpreting the usual possible worlds
definitions in an intuitionistic meta-theory. As we will see in the
following paragraphs, ideas from \cite{brauner-depaiva2006} were also
used, where the framework IHL, for \emph{intuitionistic hybrid
  logics}, is introduced. \ialc\ concepts are described as:

\[
C, D ::= A \mid \bot \mid \top
	   \mid \neg C
           \mid C \dland D 
           \mid C \dlor  D 
           \mid C \subs D 
           \mid \exists R.C 
           \mid \forall R.C 
\]
% F ::= & C \subs D \mid a\colon C \mid a R b \\
where $C,D$ stands for concepts, $A$ for an atomic concept, $R$ for an
atomic role.
% , $F$ for a formula and $a,b$ for individuals (in hybrid logic
% reading, nominals).
We could have used distinct symbols for subsumption of concepts and
the subsumption concept constructor but this would blow-up the
calculus presentation. This syntax is more general than standard \alc\
since it includes subsumption $\subs$ as a concept-forming
operator. We have no use for nested subsumptions, but they do make the
system easier to define, so we keep the general rules. Negation could
be defined via subsumption, that is, $\neg C=C\subs\bot$, but we find
it convenient to keep it in the language.  The constant $\top$ could
also be omitted since it can be represented as $\neg\bot$.

A constructive interpretation of \ialc\ is a structure $\mathcal{I}$
consisting of a non-empty set $\Delta^{\mathcal{I}}$ of entities in
which each entity represents a partially defined individual; a
refinement pre-ordering $\preceq^{\cal I}$ on $\Delta^{\cal I}$, i.e.,
a reflexive and transitive relation; and an interpretation function
$\cdot^{\cal I}$ mapping each role name $R$ to a binary relation
$R^{\cal I}\subseteq\Delta^{\cal I} \times \Delta^{\cal I}$ and atomic
concept $A$ to a set $A^{\cal I}\subseteq\Delta^{\cal I}$ which is
closed under refinement, i.e., $x \in A^{\cal I}$ and $x\preceq^{\cal
  I}y$ implies $y\in A^{\cal I}$. The interpretation $\cal I$ is
lifted from atomic concepts to arbitrary concepts via:
\[
\begin{array}{rl}
\top^{\cal I}  & =_{df} \Delta^{\cal I} \\
\bot^{\cal I}  & =_{df} \emptyset \\
(\neg C)^{\cal I } & =_{df} \{x \mid \forall y \in \Delta^{\cal I} . x \preceq y\Rightarrow  y \not\in C^{\cal I}  \} \\
(C\dland  D)^{\cal I } & =_{df} C^{\cal I}\cap  D^{\cal I} \\
(C \dlor D)^{\cal I } & =_{df} C^{\cal I} \cup D^{\cal I } \\
(C \subs D)^{\cal I } & =_{df } \{x \mid \forall y \in  \Delta^{\cal I} . 
(x \preceq y  \text{ and }  y \in C^{\cal I }) \Rightarrow y \in D^{\cal I}  \} \\
(\exists R.C)^{\cal I} & =_{df} \{ x \mid \exists y \in \Delta^{\cal I} . (x,y) \in  R^{\cal I} \mbox{ and } y \in C^{\cal I} \} \\
(\forall R.C)^{\cal I}  & =_{df} \{ x \mid \forall y \in  \Delta^{\cal I} . x \preceq y 
  \Rightarrow \forall z \in \Delta^{\cal I}. (y, z ) \in R^{\cal I} \Rightarrow z\in C^{\cal I}  \} \\
\end{array}
\]

Following the semantics of IK, the structures $\cal I$ are models for
\ialc\ if they satisfy two frame conditions:
\begin{description}
\item[F1] if $w \leq w'$ and $wRv$ then $\exists v' . w'Rv'$ and $v \leq v'$
\item[F2] if $v \leq v'$ and $wRv$ then $\exists w' . w'Rv'$ and $w \leq w'$
\end{description}

The above conditions are diagrammatically expressed as:
\[
\begin{array}{ccc}
\xymatrix{
w^{\prime}\ar@{.>}[r]^{R} \ar@{}[dr]|{(F1)} & v^{\prime} \\
w \ar[r]^{R} \ar[u]^{\le} & v\ar@{.>}[u]_{\le}
}
& \mbox{and} &
\xymatrix{
w^{\prime}\ar@{.>}[r]^{R} \ar@{}[dr]|{(F2)} & v^{\prime} \\
w \ar[r]^{R} \ar@{.>}[u]^{\le} & v\ar[u]_{\le}
}
\end{array}
\]

Our setting simplifies \cite{mendler-scheele}, since $i\mathcal{ALC}$
satisfies (like classical $\mathcal{ALC}$) $\exists R.\bot=\bot$ and
$\exists R.(C\dlor D) = \exists R.C\dlor \exists R.D$.

 In contrast
with the above mentioned approaches, ours assign a truth values to
some formulas, also called assertions, they are not concepts as in
\cite{brauner-depaiva2006}, for example.  Below we define the syntax
of general assertions ($A$) and nominal assertions ($N$) for ABOX
reasoning in \ialc. Formulas ($F$) also includes subsumption of
concepts interpreted as propositional statements.
\[
N ::=  x\colon C \mid x\colon N
\hspace{2cm}
A ::=  N \mid x R y \hspace{2cm}  F ::= A \mid C \subs C 
\]
where $x$ and $y$ are nominals, $R$ is a role symbol and $C$ is a
concept. In particular, this allows $x\colon (y\colon C)$, which is a
perfectly valid nominal assertion.

\begin{definition}[outer nominal]\label{def:outer}
  In a nominal assertion $x\colon\gamma$, $x$ is said to be the outer
  nominal of this assertion. That is, in an assertion of the form
  $x\colon(y\colon\gamma)$, $x$ is the outer nominal.
\end{definition}

% Semantic satisfaction relation $\models$ is defined in the usual
% fashion. That is,
We write $\mathcal{I},w \models C$ to abbreviate $w \in C^\mathcal{I}$
which means that entity $w$ satisfies concept $C$ in the
interpretation $\cal I$\footnote{In IHL, this $w$ is a world and this
  satisfaction relation is possible world semantics}. Further, $\cal
I$ is a model of $C$, written $\mathcal{I} \models C$ iff $\forall w
\in {\cal I}$, we have that ${\cal I}, w \models C$. Finally, $\models C$ means
$\forall {\cal I}$, we have that ${\cal I} \models C$. All previous notions are
extended to sets $\Phi$ of concepts in the usual universal
fashion. Given the hybrid satisfaction statements, the interpretation
and semantic satisfaction relation are extended in the expected
way. The statement $\mathcal{I},w\models x\colon C$ holds, if and only
if, $\forall z_{x}\succeq^{\cal I} x\ $, we have that, $ \mathcal{I},z_{x} \models C$.
  % where $z_{x}^{\mathcal{I}}$ denotes the world pointed out by $z_x$
In a similar fashion, $\mathcal{I}, w \models xRy$ holds ,if and only
if, $\forall z_{x}\succeq x.\forall z_{y}\succeq
y.(x_{x}^{\mathcal{I}},z_{y}^{\mathcal{I}}) \in R^{\mathcal{I}}$. That
is, the evaluation of the hybrid formulas does not take into account
only the world $w$, but it has to be monotonically preserved. It can
be observed that for every $w^{\prime}$, if $x^{\mathcal{I}}\preceq
w^{\prime}$ and $\mathcal{I},x^{\prime}\models \alpha$, then
$\mathcal{I},w^{\prime}\models \alpha$ is a property holding on this
satisfaction relation.

In common reasoning tasks the interpretation $\cal I$ and the entity
$w$ in a verification goal such as ${\cal I}, w \models\delta$ are not
given directly but are themselves axiomatized by sets of concepts and
formulas. Usually we have a set $\Theta$~\footnote{Here we consider
  only acycled TBox with $\subs$ and $\equiv$.} of formulas and the
set $\Gamma$ of concepts. Accordingly:

\begin{definition}\label{models}
We write $\Theta,\Gamma\models\delta$ if it is the case
that:
\begin{multline}
\forall {\cal I}.\ if\ (\forall x \in \Delta^{\cal I}.({\cal I}, x \models\Theta)  \\ 
\ then\  \forall(Nom(\Gamma,\delta)).\forall \vec{z}\succeq Nom(\Gamma,\delta).( {\cal I}, \vec{z} \models \Gamma 
\Rightarrow {\cal I},\vec{z} \models \delta)
\end{multline}
where $\vec{z}$ denotes a vector of variables $z_1,\ldots,z_k$ and
$Nom(\Gamma,\delta)$ is the vector of all outer nominals occurring in
each nominal assertion of $\Gamma\cup\{\delta\}$. $x$ is the only
outer nominal of a nominal assertion $\{x\colon \gamma\}$, while a
(pure) concept $\gamma$ has no outer nominal.
\end{definition}

A Hilbert calculus for \ialc\ is provided following
\cite{plotkin1986,simpson95,Fisher1984}. It consists of all axioms of
intuitionistic propositional logic plus the axioms and rules displayed
in Figure~\ref{fig:axioms}. The Hilbert calculus implements
TBox-reasoning. That is, it decides the semantical relationship
$\Theta, \emptyset \models C$. $\Theta$ has only formulas as members.

\begin{figure}[htbp]
\begin{align*}
0. & \quad \mbox{all substitution instances of theorems of IPL}      \\
1. & \quad \forall R.(C\subs D)\subs (\forall R.C\subs \forall R.D)  \\
2. & \quad \exists R.(C\subs D)\subs (\exists R.C\subs \exists R.D)  \\
3. & \quad \exists R.(C\dlor D)\subs (\exists R.C\dlor \exists R.D)  \\
4. & \quad \exists R.\bot\subs \bot                                  \\
5. & \quad (\exists R.C\subs \forall R.C)\subs \forall R.(C\subs D)  \\
\mathsf{MP}  & \quad \mbox{If $C$ and $C\subs D$ are theorems, $D$ is a theorem too.} \\
\mathsf{Nec} & \quad \mbox{If $C$ is a theorem then $\forall R.C$ is a theorem too.} 
\end{align*}
\caption{The \ialc\ axiomatization}\label{fig:axioms}
\end{figure}

A Sequent Calculus for \ialc\ is also provided. The logical rules of
the Sequent Calculus for \ialc\ are presented in
Figure~\ref{fig:sc-ialc}.~\footnote{The reader may want to read Proof
  Theory books, for example,
  \cite{takeuti2013,buss1998,negri2008,girard1989}.} The structural
rules and the cut rule are omitted but they are as usual. The $\delta$
stands for concepts or assertions ($x\colon C$ or $xRy$), $\alpha$ and
$\beta$ for concept and $R$ for role. $\Delta$ is a set of
formulas. In rules p-$\exists$ and p-$\forall$, the syntax $\forall
R.\Delta$ means $\{\forall R.\alpha \mid \alpha\in concepts(\Delta)
\}$,
% that $\forall_{\delta\in concepts(\Delta)} R.\delta$ 
that is, all concepts in $\Delta$ are universal quantified with the
same role. The assertions in $\Delta$ are kept unmodified. In the same
way, in rule p-N the addition of the nominal is made only in the
concepts of $\Delta$ (and in $\delta$ if that is a concept) keeping
the assertions unmodified.

The propositional connectives ($\dland,\dlor,\subs$) rules are as
usual, the rule $\dlor_2$-r is omitted. The rules are presented
without nominals but for each of these rules there is a counterpart
with nominals. For example, the rule $\subs$-r has one similar:
\begin{prooftree}
\Axiom$ \Delta, x\colon\alpha \fCenter x\colon\beta $ \RightLabel{ n-$\subs$-r }
\UnaryInf$ \Delta \fCenter x\colon(\alpha\subs\beta)$
\end{prooftree}

The main modification comes for the modal rules, which are now role
quantification rules. We must keep the intuitionistic constraints for
modal operators. Rule~$\exists$-l has the usual condition that $y$ is
not in the conclusion. Concerning the usual condition on the
$\forall$-r rule, it is not the case in this system, for the
interpretation of the a nominal assertion in a sequent is already
implicitly universal (Definition~\ref{models}).

\begin{figure}[htbp]
\begin{center}
\begin{dedsystem}
\hline
\AxiomC{}
\UnaryInf$\Delta, \delta \fCenter \delta $
\DisplayProof & & 
\AxiomC{}
\UnaryInf$\Delta, x\colon\bot \fCenter \delta $
\DisplayProof  \\
\Axiom$ \Delta, xRy \fCenter y\colon\alpha$  
\RightLabel{$\forall$-r}
\UnaryInf$ \Delta \fCenter x\colon\forall R.\alpha$
\DisplayProof & &
\Axiom$ \Delta, x\colon\forall R.\alpha, y\colon\alpha, xRy \fCenter \delta$ 
\RightLabel{$\forall$-l}
\UnaryInf$\Delta, x\colon\forall R.\alpha, xRy \fCenter \delta$
\DisplayProof \\ 
\Axiom$\Delta \fCenter xRy $
\Axiom$\Delta \fCenter y\colon\alpha$ 
\RightLabel{$\exists$-r}
\BinaryInf$\Delta \fCenter x\colon\exists R.\alpha$
\DisplayProof & & 
  \Axiom$ \Delta, xRy, y\colon\alpha \fCenter \delta $ 
\RightLabel{$\exists$-l}
\UnaryInf$  \Delta, x\colon\exists R.\alpha \fCenter \delta $
\DisplayProof \\
    \Axiom$ \Delta, \alpha \fCenter \beta $ \RightLabel{ $\subs$-r }
 \UnaryInf$ \Delta \fCenter \alpha \subs \beta $
 \DisplayProof  & &
     \Axiom$ \Delta_1 \fCenter \alpha $
     \Axiom$ \Delta_2, \beta \fCenter \delta $ \RightLabel{ $\subs$-l }
 \BinaryInf$ \Delta_1, \Delta_2, \alpha \subs \beta \fCenter \delta $
 \DisplayProof \\
     \Axiom$ \Delta \fCenter \alpha $
     \Axiom$ \Delta \fCenter \beta  $ \RightLabel{ $\dland$-r }
 \BinaryInf$ \Delta \fCenter \alpha \sqcap \beta  $
 \DisplayProof & & 
    \Axiom$  \Delta, \alpha, \beta \fCenter \delta $ \RightLabel{ $\dland$-l }
 \UnaryInf$  \Delta, \alpha \sqcap \beta \fCenter \delta $
 \DisplayProof \\
    \Axiom$  \Delta \fCenter \alpha $ \RightLabel{ $\dlor_1$-r }
 \UnaryInf$ \Delta \fCenter \alpha\sqcup\beta $
 \DisplayProof & & 
     \Axiom$ \Delta, \alpha \fCenter \delta $
     \Axiom$ \Delta, \beta  \fCenter \delta  $ \RightLabel{ $\dlor$-l }
 \BinaryInf$ \Delta, \alpha\sqcup\beta \fCenter \delta  $
 \DisplayProof \\
   \Axiom$ \Delta, \alpha \fCenter \beta $ \RightLabel{ p-$\exists$ }
\UnaryInf$ \forall R.\Delta, \exists R.\alpha \fCenter \exists R.\beta $
\DisplayProof && 
   \Axiom$ \Delta \fCenter \alpha $\RightLabel{ p-$\forall$ } 
\UnaryInf$ \forall R.\Delta \fCenter \forall R.\alpha $
\DisplayProof \\ 
   \Axiom$ \Delta \fCenter \delta $ \RightLabel{p-N}
\UnaryInf$ x\colon\Delta \fCenter x\colon\delta $
\DisplayProof & \\[3ex]\hline
\end{dedsystem}
\caption{The System \system{SC}{iALC}: logical rules}\label{fig:sc-ialc}
\end{center}
\end{figure}

\begin{theorem} 
  The sequent calculus described in Fig.~\ref{fig:sc-ialc} is sound
  and complete for TBox reasoning, that is $\Theta,\emptyset\models C$
  if and only if $\Theta\Rightarrow C$ is derivable with the rules of
  Figure~\ref{fig:sc-ialc}.
\end{theorem}

The completeness of our system is proved relative to the
axiomatization of \ialc, shown in Figure~\ref{fig:axioms}. The proof
is presented in Section~\ref{sec:completeness}.

The soundness of the system is proved directly from the semantics of
\ialc\ including the ABOX, that is, including nominals. The semantics
of a sequent is defined by the satisfaction relation, as shown in
Definition~\ref{models}. The sequent $\Theta,\Gamma \SEQ\ \delta$ is
valid if and only if $\Theta, \Gamma \models \gamma$. Soundness is
proved by showing that each sequent rule preserves the validity of the
sequent and that the initial sequent is valid. This proof is presented
in Section~\ref{sec:soundness}.

We note that although we have here fixed some inaccuracies in the
presentation of the \ialc\ semantics in \cite{PHR:2010}, the system
presented here is basically the same, excepted that here the
propositional rules are presented without nominals. Given that, the
soundness of the system proved in \cite{PHR:2010} can be still
considered valid without further problems. Note also that the proof of
soundness provides in Section~\ref{sec:soundness} is regarded the full
language of \ialc. It considers nominals and assertion on nominals
relationship, that is it concerns ABOX and TBOX. The proof of
completeness is for the TBOX only. A proof of completeness for ABOX
can be done by the method of canonical models. For the purposes of
this article, we choose to show the relative completeness proof with
the sake of showing a simpler proof concerning TBOX.

%%% Local Variables: 
%%% mode: latex
%%% TeX-master: "CooriALC"
%%% End: 

\section{The completeness of \system{SC}{iALC} system}
\label{sec:completeness}

We show the relative completeness of \system{SC}{iALC} regarding the
axiomatic presentation of \ialc\ presented in
Figure~\ref{fig:axioms}. To prove the completeness of
\system{SC}{iALC} it is sufficient to derive in \system{SC}{iALC} the
axioms 1--5 of \ialc. It is clear that all substitution instances of
IPL theorems can also be proved in \system{SC}{iALC} using only
propositional rules. The MP rule is a derived rule from the
\system{SC}{iALC} using the cut rule. The Nec rule is the p-$\forall$
rule in the system with $\Delta$ empty. In the first two proofs below
do not use nominals for given better intuition of the reader about the
use of rules with and without nominals.

\vspace{.5cm}

\noindent Axiom 1:
\begin{prooftree}
\def\fCenter{ \SEQ\ }
\Axiom$ \alpha \fCenter \alpha $
\Axiom$ \beta \fCenter \beta $ \RightLabel{$\subs$-l}
\BinaryInf$ \alpha \subs \beta, \alpha \fCenter \beta $ \RightLabel{p-$\exists$}
\UnaryInf$ \forall R.(\alpha \subs \beta), \exists R.\alpha \fCenter  \exists R.\beta $ \RightLabel{$\subs$-r}
\UnaryInf$ \forall R.(\alpha \subs \beta) \fCenter \exists R.\alpha \subs \exists R.\beta $ 
\end{prooftree}

\noindent Axiom 2:
\begin{prooftree}
\def\fCenter{ \SEQ\ }
\Axiom$ \alpha \fCenter \alpha $
\Axiom$ \beta \fCenter \beta $ \RightLabel{$\subs$-l}
\BinaryInf$ \alpha \subs \beta, \alpha \fCenter \beta $ \RightLabel{p-$\forall$}
\UnaryInf$ \forall R.(\alpha \subs \beta), \forall R.\alpha \fCenter  \forall R.\beta $ \RightLabel{$\subs$-r}
\UnaryInf$ \forall R.(\alpha \subs \beta) \fCenter \forall R.\alpha \subs \forall R.\beta $ 
\end{prooftree}

\noindent Axiom 3:
\begin{prooftree}
\def\fCenter{ \SEQ\ }
\Axiom$ xRy, y:\bot \fCenter x:\bot $ \RightLabel{$\exists$-l}
\UnaryInf$ x:\exists R.\bot \fCenter x:\bot $ \RightLabel{$\subs$-r}
\UnaryInf$ \fCenter x:(\exists R.\bot \subs \bot) $
\end{prooftree}

\noindent Axiom 4:
\begin{prooftree}\small
\def\fCenter{ \SEQ\ }
   \Axiom$ x:\exists R.\alpha \fCenter x:\exists R.\alpha $ 
\RightLabel{$\dlor_1$-r} \UnaryInf$ x:\exists R.\alpha \fCenter x:(\exists R.\alpha \dlor \exists R.\beta) $ 

   \Axiom$ x:\exists R.\beta \fCenter x:\exists R.\beta $ 
\RightLabel{$\dlor_2$-r} \UnaryInf$ x:\exists R.\beta \fCenter x:(\exists R.\alpha \dlor \exists R.\beta) $ 
\RightLabel{$\dlor$-l}   \BinaryInf$ x:\exists R.(\alpha \dlor \beta) \fCenter x:(\exists R.\alpha \dlor \exists R.\beta) $ 
\end{prooftree}

\noindent Axiom 5:
\begin{prooftree}\small
\def\fCenter{ \SEQ\ }
\AxiomC{$ xRy, y:\alpha \fCenter y:\alpha $}
\AxiomC{$ xRy, y:\alpha \fCenter xRy $}
\RightLabel{$\exists$-r} 
\BinaryInfC{$ xRy, y:\alpha \fCenter x:\exists R.\alpha $}
\AxiomC{$ xRy, y:\alpha, y:\beta, \forall R.\beta \fCenter y:\beta $}
\RightLabel{$\forall$-l} \UnaryInfC{$ xRy, y:\alpha, x:\forall R.\beta \fCenter y:\beta$}
\RightLabel{$\subs$-l}  \BinaryInfC{$ x:(\exists R.\alpha \subs \forall R.\beta), xRy, y:\alpha \fCenter y:\beta$}
\RightLabel{$\forall$-r} \UnaryInfC{$ x:(\exists R.\alpha \subs \forall R.\beta), xRy \fCenter y:(\alpha \subs \beta)$} 
\RightLabel{$\forall$-r} \UnaryInfC{$ x:(\exists R.\alpha \subs \forall R.\beta) \fCenter x:\forall R.(\alpha \subs \beta)$}
\RightLabel{$\subs$-r}   \UnaryInfC{$ \fCenter x:[(\exists R.\alpha \subs \forall R.\beta) \subs \forall R.(\alpha \subs \beta)]$}
\end{prooftree}

%%% Local Variables: 
%%% mode: latex
%%% TeX-master: "CooriALC"
%%% End: 

\section{Soundness of \system{SC}{iALC} system}
\label{sec:soundness}

In this section we prove that.

\begin{proposition} 
  If $\Theta,\Gamma \SEQ\ \delta$ is provable in \system{SC}{iALC}
  then $\Theta, \Gamma \models \gamma$.
\end{proposition}

{\em Proof}: We prove that each sequent rule preserves the validity of
the sequent and that the initial sequents are valid. The definition of
a valid sequent ($\Theta,\Gamma\models\gamma$) is presented in
Definition~\ref{models}.

The validity of the axioms is trivial. We first observe that any
application of the rules $\subs$-r, $\subs$-l,$\dland$-r,$\dland$-l,
$\dlor_1$-r,$\dlor_2$-r, $\dlor$-l of \system{SC}{iALC} where the
sequents do not have any nominal, neither in $\Theta$ nor in $\Gamma$,
is sound regarded intuitionistic propositional logic kripke semantics,
to which the validity definition above collapses whenever there is no
nominal in the sequents. Thus, in this proof we concentrate in the
case where there are nominals. We first observe that the nominal
version of $\subs$-r, the validity of the premises includes
\[
\forall(Nom(\Gamma,\delta)).\forall \vec{z}\succeq Nom(\Gamma,\delta).( {\cal I}, \vec{z} \models 
\Gamma \Rightarrow {\cal I},\vec{z} \models \delta)
\]
This means that $\Gamma$ holds in any worlds $\vec{z}\succeq\vec{x}$
for the vector $\vec{x}$ of nominals occurring in $\Gamma$. This
includes the outer nominal $x_i$ in $\delta$ (if any). In this case
the semantics of $\subs$ is preserved, since $\vec{z}$ includes
$z_i\succeq x_i$. With the sake of a more detailed analysis, we
consider the following instance:

\begin{prooftree}
\def\fCenter{ \SEQ\ }
\Axiom$ x:\alpha_1, y:\alpha_2 \fCenter x:\beta $ \RightLabel{ $\subs$-r }
\UnaryInf$ \alpha_1 \fCenter x:\alpha_2 \subs \beta $
\end{prooftree}

Consider an \ialc~structure ${\cal I}=\langle{\cal U},\preceq,R^{\cal
  I}\ldots,C^{\cal I}\rangle$ In this case, for any ${\cal I}$ and any
$z_1,z_2\in{\cal U}^{\cal I}$ if $z_1\succeq x^{\cal I}$, $z_1\succeq
y^{\cal I}$, such that, ${\cal I}, z_{i} \models \alpha_1$ and ${\cal
  I}, z_{i} \models \alpha_2$, we have that ${\cal I},z_i \models
x:\beta$, since the premise is valid, by hypothesis. In this case, by
the semantics of $\subs$ we have ${\cal I}, z_i\models
x:\alpha_1\subs\beta$. The conclusion of the rule is valid too.

The argument shown above for the $\subs$-r rule is analogous for the
nominal versions of $\subs$-r, $\subs$-l,$\dland$-r,$\dland$-l,
$\dlor_1$-r,$\dlor_2$-r, $\dlor$-l. Consider the rule $\forall$-r. 

\begin{prooftree}
\def\fCenter{ \SEQ\ }
\Axiom$ \Delta, xRy \fCenter y\colon\alpha$  
\RightLabel{$\forall$-r}
\UnaryInf$ \Delta \fCenter x\colon\forall R.\alpha$
\end{prooftree}

Since the premise is valid we have that if $\forall z_{x}\succeq
x^{\cal I}$, $\forall z_{y}\succeq y^{\cal I}$, $(z_{x},z_{y})\in
R^{\cal I}$ then $\forall z_{y}\succeq y^{\cal I}.{\cal
  I},z_{y}\models \gamma$. This entails that $x^{\cal I}\in (\forall
R.\gamma)^{\cal I}$, for $x^{\cal I}\succeq x^{\cal I}$. We observe
that by the restriction on the rule application, $y$ does not occur in
$\Delta$, it only occurs in $xRy$ and $y:\alpha$. The truth of these
formulas are subsumed by $\forall R.\gamma$. The conclusion does not
need to consider them any more. The conclusion is valid too. Another
way to see its soundness is to prove that if $xRy\Rightarrow y:\alpha$
is valid, then so is $\Rightarrow x:\forall R.\alpha$. This can be
show by the following reasoning:
\[
\forall x^{\cal I}\forall y^{\cal I}\forall z_{x}\forall
z_{y}(z_{x}\succeq x^{\cal I}\rightarrow (z_{y}\succeq y^{\cal
  I}\rightarrow ((z_{x},z_{y})\in R^{\cal I}\rightarrow {\cal
  I},z_y\models y:\alpha)))
\]
that is the same as:
\[
\forall x^{\cal I}\forall y^{\cal I}\forall z_{x}\forall
z_{y}(z_{x}\succeq x^{\cal I}\rightarrow (z_{y}\succeq y^{\cal
  I}\rightarrow ((z_{x},z_{y})\in R^{\cal I}\rightarrow {\cal
  I},y^{\cal I}\models\alpha)))
\]
Using the fact that $\forall y^{\cal I}(y^{\cal I}\succeq y^{\cal I})$,  we obtain:
\[
\forall x^{\cal I}\forall z_{x}(z_{x}\succeq x^{\cal I}\rightarrow
\forall y^{\cal I}((z_{x},y^{\cal I})\in R^{\cal I}\rightarrow {\cal
  I},y^{\cal I}\models\alpha))
\]
The above condition states that $\Rightarrow x:\forall R.\alpha$ is valid. 
\[
\forall x^{\cal I}\forall y^{\cal I}\forall z_{x}\forall
z_{y}(z_{x}\succeq x^{\cal I}\rightarrow (z_{y}\succeq y^{\cal
  I}\rightarrow ((z_{x},z_{y})\in R^{\cal I}\rightarrow {\cal
  I},z_y\models y:\alpha)))
\]

Consider the rule $\forall$-l:

\begin{prooftree}
\def\fCenter{ \SEQ\ }
\Axiom$ \Delta, x\colon\forall R.\alpha, y\colon\alpha, xRy \fCenter \delta$ 
\RightLabel{$\forall$-l}
\UnaryInf$\Delta, x\colon\forall R.\alpha, xRy \fCenter \delta$
\end{prooftree}

As in the $\forall$-r case, we analyze the simplest validity
preservation: if $x:\forall R.\alpha\land xRy$ is valid, then so is
$x:\forall R.\alpha\land y:\alpha\land xRy$. The first condition is:

\begin{multline}
\forall x^{\cal I} \forall y^{\cal I} \forall z_x 
(z_{x}\succeq x^{\cal I} \rightarrow \forall z_{y}(z_{y}\succeq y^{\cal I} \rightarrow \\
(({\cal I},z_y\models x:\forall R.\alpha)\land ({\cal I},z_y\models
x:\forall R.\alpha)\land ((z_{x},z_{y})\in R^{\cal I}) \rightarrow \\
({\cal I},z_y\models y:\alpha)\land ({\cal I},z_x\models y:\alpha))))
\end{multline}

Using $z_y=y^{\cal I}$, eliminating $z_x$ from the term, and, using
the fact that ${\cal I},z_y\models y:\alpha$ is valid, iff, ${\cal
  I},y^{\cal I}\models\alpha$ , we obtain

\begin{multline}
\forall x^{\cal I}\forall y^{\cal I}\forall z_{x}(z_{x}\succeq x^{\cal
  I}\rightarrow \forall z_{y}(z_{y}\succeq y^{\cal I}\rightarrow \\
(({\cal I},z_y\models x:\forall R.\alpha)\land ({\cal I},z_y\models
x:\forall R.\alpha)\land ((z_{x},z_{y})\in R^{\cal I})\rightarrow
({\cal I},y\models\alpha))))
\end{multline}

Consider the semantics of $\exists R.\alpha$:

\[
(\exists R.\alpha)^{\cal I}=_{df}\{ x \mid \exists y \in {\cal
  U}^{\cal I} . (x,y) \in R^{\cal I} \mbox{ and } y \in \alpha^{\cal
  I} \}
\]
and the following rule:

\begin{prooftree}
\def\fCenter{ \SEQ\ }
\Axiom$\Delta \fCenter xRy $
\Axiom$\Delta \fCenter y\colon\alpha$ 
\RightLabel{$\exists$-r}
\BinaryInf$\Delta \fCenter x\colon\exists R.\alpha$
\end{prooftree}

We can see that the premises of the rule entails the conclusion.  The
premises correspond to the following conditions:
\[
\forall x^{\cal I} \forall y^{\cal I}\forall z_{x}(z_{x}\succeq
x^{\cal I}\rightarrow \forall z_{y}(z_{y}\succeq y^{\cal I}\rightarrow
((z_{x},z_{y})\in R^{\cal I})))
\]
and
\[
\forall y^{\cal I}\forall z_{y}(z_{y}\succeq y^{\cal I}\rightarrow
(({\cal I},z_y\models y:\alpha)))
\]
Instantiating in both conditions $z_{y}=y^{\cal I}$ and $z_{x}=x^{\cal
  I}$, this yields $(x^{\cal I},y^{\cal I})\in R^{\cal I}$, such that
${\cal I},y^{\cal I}\models \alpha$, so ${\cal I}, z_{x}\models
x^{\cal I}:\exists R.\alpha$. Thus, $\exists$-r is sound. The
soundness of $\exists$-l is analogous to $\forall$-l.

Finally, it is worth noting that, for each rule, we can derive the
soundness of its non-nominal version from the proof of soundness of
its nominal version. For instance, the soundness of the nominal
version of rule~$\dlor$-l depends on the diamond conditions F1 and
F2. The soundness of its non-nomimal version, is a consequence of the
soundness of the nominal version.

The rules below have their soundness proved as a consequence of the
following reasonings in first-order intuitionistic logic that are used
for deriving the semantics of the conclusions from the semantics of
the premises: 
\begin{description}
\item[(p-$\exists$)] $\forall x(A(x)\land B(x)\rightarrow C(x))\models
  \forall x A(x)\land\exists xB(x)\rightarrow \exists xC(x)$;

\item[(p-$\forall$)] $(A(x)\models B(x))$ implies $\forall
  y(R(y,x)\rightarrow A(x))\models\forall y(R(y,x)\rightarrow B(x))$;

\item[(p-N)] if $A\models B$ then for every Kripke model ${\cal
    I}$ and world $x^{\cal I}$, if ${\cal I},x^{\cal I}\models A$ then
  ${\cal I}, x^{\cal I}\models B$.
\end{description}

\vspace{0.5cm}
\noindent \begin{tabularx}{\textwidth}{ccc}
   \Axiom$ \Delta, \alpha \fCenter \beta $ \RightLabel{p-$\exists$}
\UnaryInf$ \forall R.\Delta, \exists R.\alpha \fCenter \exists R.\beta $
\DisplayProof & 
   \Axiom$ \Delta \fCenter \alpha $\RightLabel{ p-$\forall$} 
\UnaryInf$ \forall R.\Delta \fCenter \forall R.\alpha $
\DisplayProof & 
   \Axiom$ \Delta \fCenter \delta $ \RightLabel{p-N}
\UnaryInf$ x\colon\Delta \fCenter x\colon\delta $
\DisplayProof \\
\end{tabularx}

%%% Local Variables: 
%%% mode: latex
%%% TeX-master: "CooriALC"
%%% End: 

\section[Solving Contrary-to-Duty Paradoxes in iALC]{Solving Chisholm and other contrary-to-duty paradoxes in iALC}

The paradoxes discussed in this section are known from the literature as contrary-to-duty paradoxes. They are deontic paradoxes under SDL formalization. Usually, there is a primary norm/law/obligation and a secondary norm that comes to effect when the primary obligation is violated. The form of these normative and intuitively coherent situations are in general hard to find a consistent deontic formalization. Because of that they  are called paradoxes.  A typical example of contrary-to-duty  paradox appeared in~\cite{Chisholm1963}: 

  \begin{enumerate}
  \item\label{l1} It ought to be that Jones goes to the assistance of his
    neighbors.
  \item\label{l2} It ought to be that if Jones does go then he tells them he is
    coming.
  \item\label{l3} If Jones doesn't go, then he ought not tell them he is coming.
  \item\label{fact} Jones doesn't go.
  \end{enumerate}

  \begin{itemize}
  \item This certainly appears to describe a possible situation. 1-4
    constitute a mutually consistent and logically independent set of
    sentences.

  \item (1) is a primary obligation, what Jones ought to do
    unconditionally. (2) is a \emph{compatible-with-duty} obligation,
    appearing to say (in the context of 1) what else Jones ought to do
    on the condition that Jones fulfills his primary obligation. (3)
    is a \emph{contrary-to-duty} obligation (\emph{CTD}) appearing to
    say (in the context of 1) what Jones ought to do conditional on
    his violating his primary obligation. (4) is a factual claim,
    which conjoined with (1), implies that Jones violates his primary
    obligation.

  \end{itemize}

We firstly remember the deontic approch to law and its logic. Differently of ours, it takes {\em laws} as propositions. Thus, a norm or {\em law} is an obligatory proposition, such as ``You must pay your debits'' or ``It is obligatory to pay the debits''. As a proposition each norm has a truth value. The underlying logic classical. If $\phi$ is a proposition then $O\phi$ is a proposition too. $O\phi$ intuitively means $\phi$ must be the case ,or {\em It is obligatory that $\phi$}. The paradoxes that we discuss in this work appear just when {\em laws} are taken as propositions. They show them up from the most basic deontic logic 
Standard Deontic Logic (SDL). SDL is a modal logic defined by von Wright19951~\cite{vonWright1951} and, according to the modal logic terminology on the names of axioms, it is defined by the following set of axioms. The formulas of SDL include the modality $O$.

  \begin{description}
  \item[TAUT] all tautologies of the language. This means that if $\phi$ is a propositional tautology then the substitution of $p$ for any SDL formula is an SDL tautology too;
  \item[OB-K] $O(p \to q) \to (Op \to Oq)$
  \item[OB-D] $Op \to \neg O\neg p$
  \item[MP] if $\vdash p$ and $\vdash p \to q$ then $\vdash q$
  \item[OB-NEC] if $\vdash p$ then $\vdash Op$ 
  \end{description}

  SDL is just the normal modal logic \emph{D} or \emph{KD}, with a
  suggestive notation expressing the intended interpretation. 
  From these, we can prove the principle that obligations cannot
  conflict, \emph{NC of SDL}, $\neg(Op \land O\neg p)$, see~\cite{vonWright1951}.

The following set of formulas is a straightforward formalization of Chisholm paradox in SDL. 

  \begin{enumerate}
  \item $Op$
  \item $O(p \to q)$
  \item $\neg p \to O\neg q$
  \item $\neg p$
  \end{enumerate}

The intuitive meaning of each formula is according the table shown in figure~\ref{chisholm} where $p$ is ``Jones go to  the assistance of his neighbours'' and $q$ is ``Jones tells his neighbours he is going''. 

\begin{figure}
\begin{tabular}{|c|c|}
\hline
Assertion & SDL Formula \\
\hline
\hline
It ought to be that Jones go to & \\
  the assistance of his neighbours. & $O(p)$ \\
\hline
It ought to be that if Jones does go then & \\
 he tells them he is going. & $O(p\rightarrow q)$ \\
\hline
If Jones doesn't go, then & \\
 he ought not tell them he is going. & $\neg p\rightarrow O(\neg q)$ \\
\hline
Jones doesn't go. & $\neg p$ \\
\hline
\end{tabular}
\caption{Assertions of the Chisholm paradox and their representation in SDL}
\label{chisholm}
\end{figure}

 Using the deductive power of SDL we can perform the following derivation of a SDL contradiction. 
  \begin{itemize}
  \item from (2) by principle \emph{OB-K} we get $Op \to Oq$, 
  \item and then from (1) by \emph{MP}, we get $Oq$;
  \item but by \emph{MP} alone we get $O\neg q$ from (3) and
    (4). 
  \item From these two conclusions, by \emph{PC}, we get
    $Oq \land O\neg q$ , contradicting \emph{NC of SDL}.
  \end{itemize}

  Assertion 1-4, from Chisholm paradox,  leads to inconsistency per SDL. But, 1-4 do not seem
  inconsistent at all, the representation cannot be a faithful one. We discuss this in the sequel.
For reasons that will become clear, we take Chisholm paradox as stated above in natural language, instead of its SDL version. We use the same letters to denote the propositions/laws as used in the deontic representation of the paradox, for a better comparison.

In first law in the paradox, i.e., the law state in item~\ref{l1} is a nominal in \ialc, and hence it is a Kripke world in our model. The same can be said about item~\ref{l2}. The state-of-affairs, expressed in \ialc, is simply the assertions: $l_1:\top$ and $l_2:\top$. Note that this assertions only state that there are two laws $l_1$ and $l_2$ in the legal universe. Since a Kripke model for intuitionistic logic is a Heyting algebra, and hence it is a lattice too, there must be the {\em meet} of these two worlds.
This is represented in the model by law $l_0$, intuitively stating that it is obligatory to do what law $l_1$ and law $l_2$ state. Item~\ref{l3} of the paradox is a conditional that generally states that if some proposition is truth then some law exists. This is a rather hard expression in judicial terms. Laws exist by promulgation only, they do not have their existence conditioned to anything but their own promulgation. This conditional expression can be raised in a legislative discussion only. But even in this exceptional case, the raising of paradoxes, as the one under discussion, advices that such use should be avoided. What item~\ref{l3} says, instead, is simply that $\neg p$ holds in the world $l_3$ that is the law cited as the consequent of the conditional. Finally, as the model is a lattice, there must be a world $l_4$ that represents the law that it is the conjunctive law related to $l_0$ and $l_3$, in the same way $l_0$ is related to $l_1$ and $l_2$. Now, in $l_4$, it is ensured that $\neg p$ holds by the intuitionistic interpretation of the negation. As a result the model depicted in the diagram below is a model for what is known by Chisholm paradox. Thus, it is not a paradox when expressed in \ialc in a kelsenian way.

  \begin{enumerate}
  \item The law $l_1$, originally $Op$
  \item The law $l_2$, originally $O(p \to q)$
  \item From (3), $\neg p \to O\neg q$, we have $l_3:\neg p$. If we
    had $O\neg q \to \neg p$ the translation would be the same. That
    is, $l_3$ is $O\neg q$.
  \item The law $l_0$ that represents the infinum of $l_1$ and $l_2$.
    % $l$ is the $glb(l_1,l_2)$, the intuitionistic precedence
    % relation is a reticulate.
  \end{enumerate}

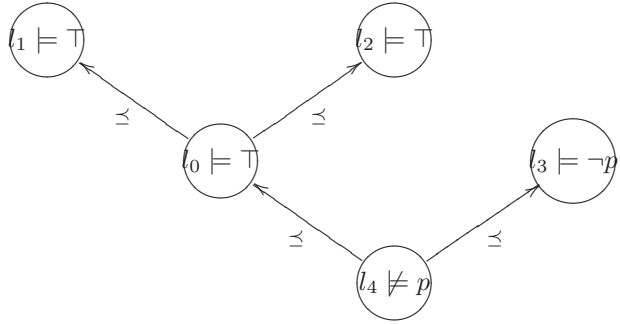
\begin{figure}
  \xymatrix{*++[o][F]{\mbox{$l_1\models\top$}} & & *++[o][F]{\mbox{$l_2\models\top$}} &  \\
    & *++[o][F]{\mbox{$l_0\models\top$}}\ar[ul]^{\preceq}\ar[ur]_{\preceq} & & 
    *++[o][F]{\mbox{$l_3\models\neg p$}}   \\
    & & *++[o][F]{\mbox{$l_4\not\models p$}}\ar[ul]^{\preceq}\ar[ur]_{\preceq} & }
\caption{Kripke model for the \ialc representation of Chisholm paradox}
\label{model}
\end{figure}

The diagram  in figure~\ref{model},  shows the Kripke model to Chisholm paradox discussed above. Remember that if $x:A$ then $\forall x' \geq x, x':A$.

Only to estimate the range of our approach for solving semantical (contrary-to-duty) paradoxes. The free choice permission paradox reported on~\cite{Ross1941}) is derived on the supposition that there is a logic of norms, as for example SDL. The dilemma on the existence of such a logic of normas is also known as Jorgensen dilemma.
At the next section we draw the conclusion that \ialc\ is more likely to such a logic than SDL. In fact in the sequel we detail the free choice permission paradox in SDL and we can see that the steps based on the $K$ axiom cannot be derived inside \ialc. 
 
 Consider the tautology $p\rightarrow p\lor q$. We have by necessity SDL rule that:
\[ 
O(p\rightarrow p\lor q)
\]
 is derivable in SDL. By the axiom $K$: 
\[
O(p)\rightarrow O(p\lor q)
\]
is also derivable.  On the other hand, by contra-positive on the first sentence,  we have
\[
 \neg (p\lor q)\rightarrow \neg p
\], so,  and hence
\[
O(\neg(p\lor q))\rightarrow O(\neg q)
\]
by $K$ necessitation and $K$ again. Thus, by contra-positive again, we obtain
\[
 \neg O(\neg q)\rightarrow \neg O(\neg (p\lor q))
\]
Taking $\neg O(\neg \phi)$ as permitted $\phi$, rather better $P(\phi)$, we draw:
\[
 P(p)\rightarrow P(p\lor q)
\]
The free choice permission states that if $P(p\lor q)$ holds then $P(p)\land P(q)$. Consider that we accept the free choice permission, so to say the formula
\[
 P(p\lor q)\rightarrow (P(p)\land P(q))
\]
As $p\rightarrow (p\lor q)$, then by what was discussed above, $P(p)\rightarrow P(p\lor q)$, then by the free choice permission, we draw that
\[
 P(p)\rightarrow P(p\land q)
\]
for any $q$. In summary, in the presence of the free choice permission axiom, we can derive that $P(p)\rightarrow P(q)$ for every $q$, which should not happen. If we use \ialc\ together with Kelsen jurisprudence, and hence intuitionistic logic, we cannot derive all the steps above, for $K$ cannot be used. Anyway, our definition of permission is rather different from what was used in this paragraph, see section~\ref{conclusion}. We can conclude that many paradoxes that are based on the axiom $K$ are not legal paradoxes in a Kelsenian formalization of legal situations in \ialc\  anymore.

\section{Conclusion}
\label{conclusion}
In this article, we shown how intuitionistic logic and Kelsen's jurisprudence can be used to express Chisholm paradox faithfully. A key fact in providing a logical model to this paradox is that laws/norms are not taken as propositions. For example, in the explanation above on building the model, if we turn back to deontic expression of laws, we will have that $l_1$ is $Op$ and $l_2$ is $O(p\rightarrow q)$, but we cannot derive that $l_3$ is $O(q)$. $l_3$ is of course the meet ($\dland$) between $l_1$ and $l_2$, as a meet it is strongly connection to $O(l_1)\land O(l_2)\leftrightarrow O(l_1\land l_2)$, which is a SDL valid formula. Thus, $l_3$ is the norm $O(l_1\land l_2)$, that is an obligation. However, now remembering  what norms $l_1$ and $l_2$ are in this particular case, $l_3$ is the meeting $O(p)\land O(p\rightarrow q)$ that it is $O(p\land (p\rightarrow q))$. This conclusion, however, does not entail that in $l_3$ can be identified with $O(q)$, since our implication is the intuitionistic implication. This very last aspect of joining Kelsen jurisprudence and \ialc\  also helps to avoid other deontic paradoxes.  

Jorgensen's Dilemma~\cite{Jorgensen1937} offers a question, in fact, a dilemma, whether there is, in fact, any deontic logic. The question follows this path: 1) Norms/laws deal with evaluative sentences; 2) Evaluative sentences are not the kind of sentence that can be true or false; 3) Thus, how there is a logic of evaluative sentences? 4) Logic has as goal to define what can be drawn from whatever, and; 5) A sentence follows from a set of sentences on a basis of the relationship between the truth of the sentences in question. Thus, there is no deontic logic. What we have shown in this article, is that deontic logic is possible by considering the logic of norms as a logic on norms, instead. This reading is just what we do in legal ontologies.

We have to touch some aspects that are very well-known in the deontic approach. One is the deontic concept of {\em permission}. This case is modeled by observing that in a society regulated by law, permission is nothing more than an obligation of the State. The State promulgates what is allowed. Concerning prohibitions, the foundation is analogous. However, some subtle and theoretical problems may arise if one wants to recover the definition of forbidden ($F$) regarding the very well-known duality $F(p)\equiv O(\neg p)$. This discussion will be the subject of another article. 

Finally, we would like to comment that professor Farinas del Cerro taught us that the research on logic and AI, mainly the first should be approached by solving part-by-part the problem and elegantly putting everything together. Well, we learned the first part of this technique, by following him by reading his articles. We think that in the first part, so to say finding a good foundation on legal ontologies, one that comes from the domain itself, namely Kelsen jurisprudence. Concerning the second part, that is to put everything together in an elegant way, we known that we are far from it. Another thing that we might have learned from prof. Farinas is that in this case, it is a matter of time to have the work in a more mature stage. We hope we reach this stage.

%%% Local Variables:
%%% mode: latex
%%% TeX-master: "CD-Paradoxes-iALC"
%%% End:

%\bibliographystyle{apacite}
%\bibliography{CD-Paradoxes}

\end{document}